\documentclass[aps,prl,twocolumn]{revtex4}
\usepackage{graphicx}

\begin{document}

\title{Possible Anderson transition below two dimensions
in disordered systems of noninteracting electrons}

\author{Yoichi Asada$^1$, Keith Slevin$^2$, and Tomi Ohtsuki$^{3,4}$}
\affiliation{
$^1$Department of Physics, Tokyo Institute of Technology,
2-12-1 Ookayama, Meguro-ku, Tokyo 152-8551, Japan\\
$^2$Department of Physics, Graduate School of Science,
Osaka University, 1-1 Machikaneyama, Toyonaka, Osaka 560-0043, Japan\\
$^3$Department of Physics,
Sophia University, Kioi-cho 7-1, Chiyoda-ku, Tokyo 102-8554, Japan\\
$^4$ CREST, JST 
}

\date{\today}

\begin{abstract}
We investigate the possibility of an Anderson transition below two
dimensions in disordered systems of non-interacting electrons with
symplectic symmetry. Numerical analysis of energy level statistics
and conductance statistics on Sierpinski carpets with spin-orbit
coupling indicates the occurrence of an Anderson transition below two
dimensions.
\end{abstract}

\pacs{}

\maketitle


According to the scaling theory of Abrahams {\it et al.} for
non-interacting electrons in disordered systems, all states are
localized in two dimensions (2D) and the Anderson transition occurs
only above 2D \cite{abrahams:79}. This prediction applies to systems
with orthogonal symmetry, i.e., systems with time reversal and spin
rotation symmetry \cite{hikami:80,mackinnon:83,schreiber:96}.
Accordingly, it is believed that the lower critical dimension for
the Anderson transition in systems with orthogonal symmetry is
\begin{equation}
d_{\mathrm L}^{\left(\mathrm{orth}\right)}=2.
\end{equation}
The prediction of Abrahams {\it et al.} does not apply to systems
with symplectic symmetry, i.e. in systems with time reversal
symmetry but in which spin rotation symmetry is broken by spin-orbit
coupling \cite{hikami:80,ando:89}. For such systems it is known that
there is a transition in 2D, if the spin-orbit interaction is
sufficiently strong.

The prediction of Abrahams {\it et al.} rests on an argument
concerning the form of the $\beta$ function that describes the
scaling of the conductance. Reasonable assumptions concerning the
asymptotic behavior of this function in the strongly metallic and
localized limits, respectively, and the assumption that the $\beta$
function is monotonic lead to their conclusion. As we explain below
it is this latter assumption that does not hold in systems with
symplectic symmetry.

\begin{figure}[tb]
\includegraphics[width=0.95\linewidth]{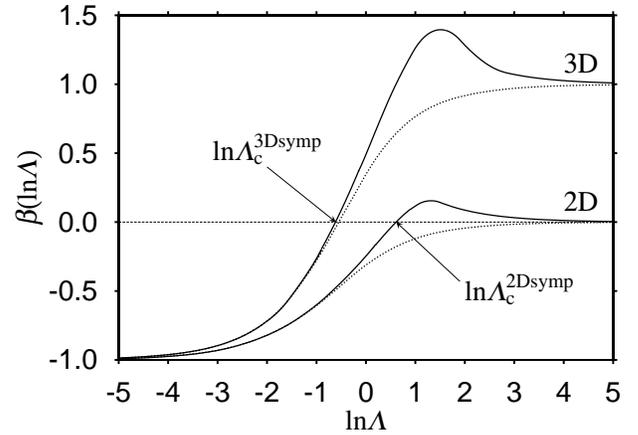}
\caption{ The numerically estimated $\beta$ functions for systems
with symplectic symmetry in 2D and 3D (solid lines)
\cite{asada:04-1,asada:05-1} as well as the schematic ones for
systems with orthogonal symmetry (dotted lines). }
\label{fig:betafunc}
\end{figure}

In Fig.~\ref{fig:betafunc} we show the $\beta$ function that
describes the scaling of the quantity $\Lambda$\cite{mackinnon:83}
\begin{equation}
 \beta(\ln \Lambda)=\frac{\mathrm{d}\ln \Lambda}{\mathrm{d}\ln L}.
 \label{eq:betafunc-lambda}
\end{equation}
Here $\Lambda$ is the ratio of the quasi-1D localization length to
the system width for electrons on a long quasi-1D system of width
$L$. The $\beta$ functions that describe the scaling of different
physical quantities are expected to differ in detail, but for
systems with the same dimensionality and symmetry all are expected
to share some common properties. For example, a zero of the $\beta$
function indicates a transition and this property should be common
to all the $\beta$ functions. Also the slope at the zero, which is
related to the critical exponent $\nu$ describing the divergence of
the localization length, should be the same for all the $\beta$
functions.

The solid lines in Fig.~\ref{fig:betafunc} are numerical estimates
of the $\beta$ function (\ref{eq:betafunc-lambda}) for systems with
symplectic symmetry \cite{asada:04-1,asada:05-1}. The $\beta$
function tends to $d-2$, with $d$ being the dimensionality, in the
metallic limit and becomes negative in the localized limit. In
contrast to systems with orthogonal symmetry (shown schematically,
see Ref.~\cite{mackinnon:83} for numerical result) the $\beta$
functions are non-monotonic and a fixed point exists even in 2D. It
is thought that the $\beta$ function for the conductance behaves
similarly \cite{wegner:89,hikami:92}.

Looking at Fig.~\ref{fig:betafunc} the obvious question is, ``What
happens below 2D for systems with symplectic symmetry?'' Does the
$\beta$ function change discontinuously and suddenly become
monotonic? Or does it change smoothly? In the latter case, we expect
the Anderson transition to persist below 2D. This is a long standing
problem \cite{hikami:85}. Our numerical results suggest that the
Anderson transition persists below 2D for systems with symplectic
symmetry. It is known that states in (quasi-)1D systems with
symplectic symmetry are always localized, so the lower critical
dimension must be at least one. This would mean that
\begin{equation}
1\le d_{\mathrm L}^{\left(\mathrm{symp}\right)}<2 .
\end{equation}

To study this problem we have simulated non-interacting electrons on
the Sierpinski carpet \cite{mandelbrot1} with spin-orbit coupling.
The Sierpinski carpet is constructed by repeating the following
iteration. We start from a $b\times b$ square lattice with a central
$c\times c$ square removed. This first generation or ``generator" is
denoted as SC($b,c,1$). At each step the next generation is
constructed by magnifying the lattice $b$ times and replacing each
site of the current generation with the generator. The linear size
of the $k$th generation SC($b,c,k$) is $L=b^k$ and the number of
sites is $N=(b^2-c^2)^k$. The Sierpinski carpet becomes a true
fractal in the mathematical sense when the generation number $k$
tends to infinity. This true fractal is denoted as SC($b,c$).
According to a lot of studies of thermal magnetic phase transitions,
the Sierpinski carpets are useful to study phase transitions in
fractal dimensionality less than two \cite{gefen:83} (see also a
recent work Ref.~\cite{monceau:01} and references therein).
We have studied SC($3,1,k$) and SC($5,1,k$) with fixed boundary
conditions imposed in both directions.

A fractal system has a fractal dimension $d_{\mathrm f}$. This
dimension describes how the mass of the fractal depends on the size
of the fractal. The fractal dimension $d_{\mathrm f}$ of the
Sierpinski carpet SC($b,c$), defined by $N=L^{d_{\mathrm f}}$, is
equal to $d_{\mathrm f}=\ln(b^2-c^2)/\ln b$. We find $d_{\mathrm
f}\approx 1.893$ for SC($3,1$) and $d_{\mathrm f}\approx 1.975$ for
SC($5,1$). A fractal system also has a spectral dimension
$d_{\mathrm s}$ \cite{nakayama}. This dimension describes the
spectra of the low energy vibrations on the fractal. Numerical
studies of the Anderson transition in bifractal systems indicate
that this is the relevant dimension when discussing the Anderson
transition \cite{schreiber:96,travenec:02}. The spectral dimension
of the Sierpinski carpet has been estimated by simulating a
classical random walk on the fractal \cite{fujiwara:95}. The  values
reported for SC($3,1$) and SC($5,1$) are, respectively, $d_{\mathrm
s}=1.785\pm 0.008$ and $d_{\mathrm s}=1.940\pm 0.009$.

We have simulated the SU(2) model \cite{asada:02-1,asada:04-1}.
\begin{equation}
 H=\sum_{i,\sigma}\epsilon_i c_{i\sigma}^{\dagger} c_{i\sigma}
- \sum_{\langle i, j\rangle,\sigma,\sigma'} R(i,j)_{\sigma\sigma'}
c_{i\sigma}^{\dagger} c_{j\sigma'}. \label{eq:H}
\end{equation}
Here $c_{i\sigma}^{\dagger}(c_{i\sigma})$ denotes the creation
(annihilation) operator of an electron at the site $i$ with spin
$\sigma$. This model has symplectic symmetry. There is an on-site
random potential $\epsilon_i$ which is identically and independently
distributed with uniform probability in the range $[-W/2,W/2]$.
There is also nearest neighbor hopping between lattice sites with
the hopping matrices $R(i,j)$ sampled uniformly from the group
SU(2). Apart from the fractal lattice the model is exactly as
described in \cite{asada:02-1,asada:04-1} and the reader should
refer to these for further details.

To investigate the localization of electrons on these fractals we
have examined the energy level statistics of (\ref{eq:H}). In
particular, we have looked at the distribution $P(s)$ of the nearest
neighbor level spacing $s$ measured in units of the mean level
spacing \cite{shklovskii:93}. It is known that $P(s)$ is well
approximated by the Wigner surmise for the Gaussian symplectic
ensemble $P_{\rm GSE}(s)=A s^4 \mathrm{e}^{-B s^2}$ in the extended
limit, and the Poisson distribution $P_{\mathrm
P}(s)=\mathrm{e}^{-s}$ in the localized limit. We have calculated
the energy eigenvalues using either LAPACK or the Lanczos method
\cite{cullum}. An unfolding procedure has been applied to the level
spacings to compensate for the variation with energy of the ensemble
average density of states. For a quantitative analysis of $P(s)$, we
define
\begin{equation}
 Y_{s_0}=\frac{\int_0^{s_0}\mathrm{d}s P(s)
-\int_0^{s_0}\mathrm{d}s P_{\mathrm{P}}(s)}
{\int_0^{s_0}\mathrm{d}s P_{\mathrm{GSE}}(s)
-\int_0^{s_0}\mathrm{d}s P_{\mathrm{P}}(s)}.
\end{equation}
In the extended limit $Y_{s_0}=1$, and in the localized limit
$Y_{s_0}=0$. Note that since each energy level has a twofold
Kramers' degeneracy we consider only distinct eigenvalues when
calculating the spacing distribution.

We have simulated lattices SC($5,1,k$) up to $k=4$. Calculation of
higher generations is not possible at present because the CPU time
required is too large. For the first
generation $k=1$ the number of eigenvalues is small. Therefore, we
restricted our analysis to a single pair of consecutive levels per
sample, the smallest positive eigenvalue and the largest negative
eigenvalue. When we choose the pair of levels in this special way,
the probability $\tilde{P}(s)$ of a level spacing with a value $s$
is equal to $sP(s)$. Therefore we multiply $\tilde{P}(s)$ by $1/s$
to obtain level spacing distribution $P(s)$. We  simulated 1000000
samples for $k=1$. For higher generations the number of eigenvalues
is greater so we reverted to the usual procedure of analyzing a
sequence of consecutive levels. The number of samples was reduced
accordingly. We used levels in the interval $E=[-0.3,0.3]$. We
simulated 15000, 600 and 10 samples for $k=$2, 3 and 4,
respectively. The total number of the level spacings for each of the
($k,W$) pair is thus of order $10^5$ to $10^6$.

\begin{figure}[tb]
\includegraphics[width=0.88\linewidth]{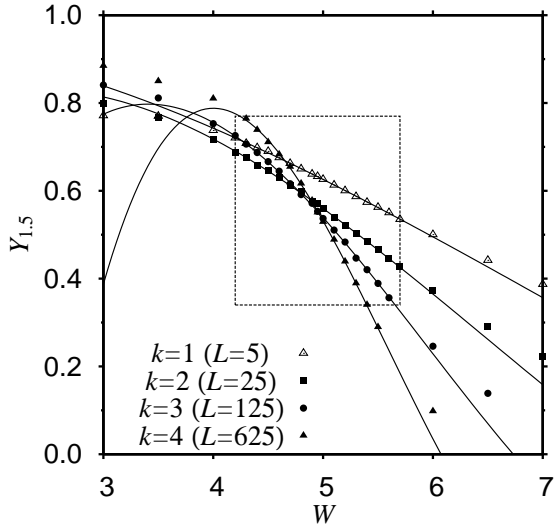}
\caption{ $Y_{1.5}$ vs. $W$ for SC($5,1,k$) ($k=1,2,3,4$). The solid
lines are the best fits of (\ref{eq:yirr}) to the data inside the
box (shown by dashed lines). Outside the box the fit deviates from
the numerical data because the expansion parameters $\psi L^{1/\nu}$
and $w$ are no longer small.} \label{fig:sc51els}
\end{figure}

\begin{table*}[tb]
\caption{\label{table:critical} The details of the scaling analyses
for SC($5,1,k$). Data with $W$ in $[4.2,5.7]$ have been used. The
range of $Y_{s_0}$ has also been restricted as indicated. Here,
$N_{\mathrm{p}}$ is the number of parameters, $N_{\mathrm{d}}$ the
number of data, $Q$ the goodness of fit probability, and
$Y_{\mathrm{c}}$ the critical value of $Y_{s_0}$. The precision of
the fitted parameters is expressed as 95\% confidence intervals.}
\begin{ruledtabular}
\begin{tabular}{ccccccccccc}
$s_0$ & $Y_{s_0}$ & $n_R,n_I,m_R,m_I$ & $N_{\mathrm{p}}$ & $N_{\mathrm{d}}$
& $Q$ & $W_{\mathrm{c}}$ & $Y_{\mathrm{c}}$ & $\nu$ & $y$  \\
\hline
0.5 & [0.60,0.93] & 4,1,2,0  & 11 & 63
& $0.3$ & $4.99\pm 0.02$  & $0.791\pm 0.009$ & $3.50\pm 0.11$
& $-0.80 \pm 0.09$ \\
1.5 & [0.34,0.77] & 3,1,2,0 & 10 & 63
& $0.7$ & $4.98\pm 0.03$ & $0.536\pm 0.013$ & $3.44\pm 0.15$
& $- 0.77\pm 0.11$
\end{tabular}
\end{ruledtabular}
\end{table*}

\begin{figure}[tb]
\includegraphics[width=0.88\linewidth]{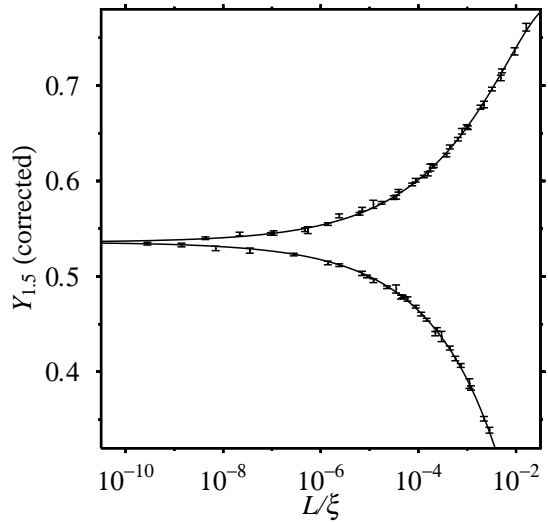}
\caption{
The numerical data $Y_{1.5}$ after subtraction
of the corrections to scaling
plotted as function of the ratio $L/\xi$.
}
\label{fig:sc51y15sps}
\end{figure}

Figure~\ref{fig:sc51els} shows $Y_{1.5}$ as a function of disorder
$W$ for SC($5,1,k$). There are large corrections to scaling in the
first generation. The data for the higher generations $k=2,3$ and
$4$ indicate that an Anderson transition occurs on SC($5,1$) around
$W\approx 5$. When disorder is weak enough, $Y_{1.5}$ increases with
$k$ for $k\geq 2$ indicating a delocalized phase. When disorder is
strong enough, $Y_{1.5}$ decreases as $k$ increases indicating a
localized phase.

Further analysis of the level statistics on SC(5,1) is based on the
assumption that finite size scaling applies to fractals in the same
way as to systems with integral dimension. Accordingly the statistic
$Y_{s_0}$ should, in the absence of any corrections to scaling, obey
the single parameter scaling law
\begin{equation}
 Y_{s_0}=f_{\pm}\left(L/\xi \right)=F_0\left(\psi L^{1/\nu}\right).
 \label{eq:ysps}
\end{equation}
Here $\xi$ is the localization(correlation) length, $\nu$ the
critical exponent for the divergence of $\xi$, and $\psi$ a smooth
function of disorder $W$ that crosses zero linearly at the critical
disorder $W_{\mathrm c}$. The subscript $\pm$ distinguishes the
scaling function in the delocalized and localized phases. It is,
however, clear from Fig.~\ref{fig:sc51els} that corrections to
scaling are not negligible. Taking account of corrections up to the
first order in an irrelevant variable $\phi L^y$ we
have \cite{slevin:99,slevin:01-1,slevin:03-1}
\begin{equation}
 Y_{s_0}=F_0 \left(\psi L^{1/\nu} \right)
+\phi L^y  F_1\left(\psi L^{1/\nu}\right).
 \label{eq:yirr}
\end{equation}
For the purpose of fitting, the scaling functions $F_0$ and $F_1$
are approximated by a power series up to the order $n_R$ and $n_I$
in $\psi L^{1/\nu}$. The functions $\psi$ and $\phi$ are also
approximated by expansions in terms of dimensionless disorder
$w=(W_{\mathrm{c}}-W)/W_{\mathrm{c}}$ up to the order $m_R$ and
$m_I$. The best fit to data is determined by minimizing the $\chi^2$
statistics and the precision of the parameters are determined using
the Monte Carlo method \cite{numrep}.

The best fit is shown in Fig.~\ref{fig:sc51els} and the results of
the scaling analyses for $s_0=0.5$ and $s_0=1.5$ are tabulated in
Table~\ref{table:critical}. The estimates of the critical disorder
$W_{\mathrm{c}}$ and the critical exponent $\nu$ for $s_0=0.5$ and
$s_0=1.5$ are consistent as required. Our estimate $\nu=3.4\pm 0.2$
for SC(5,1) is clearly different from the value $\nu=2.746\pm 0.009$
in 2D\cite{asada:04-1}, reflecting the difference of the
dimensionality. To exhibit single parameter scaling the data are
re-plotted, after subtraction of the corrections due to an
irrelevant variable, in Fig.~\ref{fig:sc51y15sps}. The two branches
correspond to the delocalized and localized phases.

\begin{figure}[tb]
\includegraphics[width=0.88\linewidth]{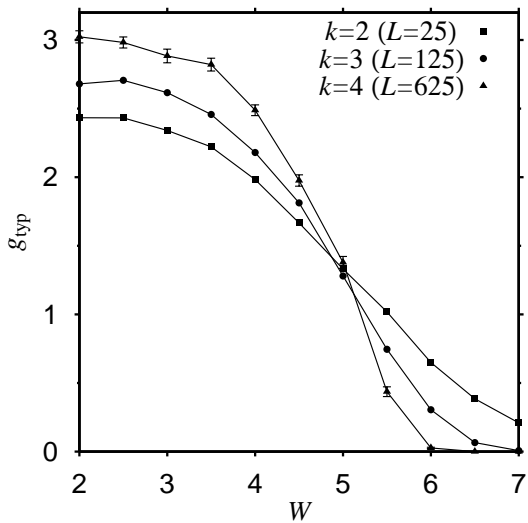}
\caption{ The typical conductance $g_{\rm typ}$ vs. $W$ for
SC(5,1,k) ($k=2,3,4$). The lines are a guide to the eye only. }
\label{fig:sc51gtyp}
\end{figure}


To complement our study of the level statistics, we have also
analyzed statistics of the Landauer conductance $g$ (in units of
$e^2/h$) \cite{shapiro:87,slevin:01-1,slevin:03-1}. Two perfect
leads of width $L=5^k$ are attached to the Sierpinski carpet
SC($5,1,k$) and the recursive Green's function method is used to
calculate the conductance \cite{ando:91}. The hopping matrices in the
direction transverse to the current are set to the unit matrix. As a
scaling quantity, we choose the typical conductance
\begin{equation}
 g_{\rm typ}=\mathrm{e}^{\langle \ln g \rangle}.
\end{equation}
Here $\langle \rangle$ means the ensemble average. We have set the
Fermi energy to $E=0$ and have accumulated 3000 samples for $k=2,3$
and from 150 to 300 samples for $k=4$. Figure~\ref{fig:sc51gtyp}
shows the typical conductance $g_{\rm typ}$ as a function of $W$ for
SC($5,1,k$). These data also indicate the occurrence of an Anderson
transition at about $W\approx 5$.

We have also analyzed energy level statistics and conductance
statistics on SC($3,1,k$) up to the fourth generation. We did not
find any evidence for a delocalized phase. We cannot conclude from
this that SC($3,1$) is below the lower critical dimension because
the SU(2) model has significant randomness even when the on-site
random potential is zero.


\begin{figure}[tb]
\includegraphics[width=0.95\linewidth]{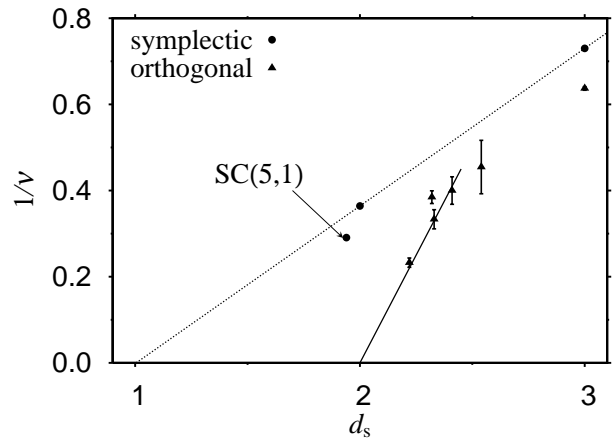}
\caption{ $1/\nu$ vs. $d_s$ for systems with orthogonal and
symplectic symmetries. The values for 2D and 3D symplectic
systems are from \cite{asada:04-1} and \cite{asada:05-1},
respectively.
The estimates for systems with orthogonal
symmetry with dimension $2<d_{\mathrm s}<3$ are from
\cite{schreiber:96} and that in 3D is from \cite{slevin:99}.
(The exponent $\nu$ for fractals with $2<d_s<3$ was also estimated
in Ref.~\cite{travenec:02}).
The solid line indicates $\nu=1/(d_{\mathrm{s}}-2)$.}
\label{fig:nuvsds}
\end{figure}

In Fig.~\ref{fig:nuvsds} we have plotted the available results for
the critical exponent for systems with orthogonal and symplectic
symmetries. For 2D and 3D system with symplectic symmetry the
critical exponent are estimated to be $\nu=2.746 \pm 0.009$ in 2D
\cite{asada:04-1} and $\nu=1.37 \pm 0.02$ in 3D \cite{asada:05-1}.
For systems with orthogonal symmetry we have used the estimates
reported in \cite{schreiber:96,slevin:99}. In the figure we plot
$1/\nu$ as a function of $d_{\mathrm{s}}$.

We recall that the slope of the $\beta$ function at the fixed point
is equal to $1/\nu$. Assuming that the $\beta$ function changes
continuously as the dimensionality changes, then just above the
lower critical dimension the fixed point should correspond to the
maximum of the $\beta$ function. We therefore expect the critical
exponent $\nu$ to diverge at the lower critical dimension. This is
consistent with the theoretical analysis of the nonlinear $\sigma$
model with orthogonal symmetry where it was found that
$\nu=1/\epsilon$ for $\epsilon=d-2 \ll 1$ \cite{hikami:81}. The
numerical data for orthogonal systems shown in Fig.~\ref{fig:nuvsds}
are consistent with this.

For systems with symplectic symmetry fewer data are available
\cite{asada:04-1,asada:05-1}. When
only the estimates for $\nu$ in 2D and 3D are considered, it can
leave the impression (cf. the dotted line in Fig.~\ref{fig:nuvsds})
that the lower critical dimension is 1 in this case. However, our
estimate of the critical exponent on SC($5,1$) is well below the
dotted line, suggesting that the lower critical dimension is closer
to 2 than to 1.

To confirm the occurrence of the Anderson transition below 2D for
systems of non-interacting electrons in disordered systems with
symplectic symmetry, and to determine more precisely the lower
critical dimension, further numerical studies on a variety of
fractals over a larger range of generations are needed. Analysis of
fractals with $d_{\mathrm s}$ intermediate between the values for
SC(3,1) and SC(5,1) might be particularly helpful. The nature of the
possible delocalized phase also warrants further study.

Recently electron transport through systems with fractal perimeter was
studied experimentally \cite{walling:04}. 
Electron transport through fractal structures can be an interesting
topic for further theoretical and experimental investigation.

One of the authors (Y.A.) is supported by Research Fellowships of the Japan
Society for the Promotion of Science for Young Scientists.


\end{document}